\documentclass[
    ,final            % use final for the camera ready runs
  ]
  {aipproc}

\layoutstyle{6x9}

\begin{document}

\title{Alkali Line Profiles in Degenerate Dwarfs}

\classification{95.30.Ky,97.10.Ex,97.20.Vs,97.20.Rp}
%<Replace this text with PACS numbers; choose from this list:
%                \texttt{http://www.aip..org/pacs/index.html}>}
\keywords      {brown dwarfs,white dwarfs,stellar atmospheres,line profiles}

\author{Derek Homeier}{
  address={Institut f{\"u}r Astrophysik, Georg-August-Universit{\"a}t,
  Friedrich-Hund-Platz 1, 37077 G{\"o}ttingen, Germany}}
\author{Nicole Allard}{ 
  address={Institut d'Astrophysique de Paris, CNRS, 
  98bis Boulevard Arago, 75014 Paris, France}}
\author{France Allard}{ 
  address={Universit\'e de Lyon, Lyon, 69003, France; Ecole Normale Sup\'erieure de
  Lyon, 46 all\'ee d'Italie, Lyon, 69007, France;
  CNRS, UMR 5574, 
  Centre de Recherche Astrophysique de Lyon; 
%%  CRAL; 
  Universit\'e Lyon~1, Villeurbanne, 69622 France}}

%\author{<author3>}{
%  address={<common address for author2 and author3>}
%  ,altaddress={<author1 address>} % additional visiting address
%}

\begin{abstract}
Ultracool stellar atmospheres show absorption by alkali resonance
lines severely broadened by collisions with neutral perturbers. In the
coolest and densest atmospheres, such as those of T dwarfs,
Na\,\textsc{i} and K\,\textsc{i} broadened by molecular hydrogen and
helium can come to dominate the entire optical spectrum. 
Their profiles have been successfully modelled with accurate
interaction potentials in the adiabatic theory, computing line
profiles from the first few orders of a density expansion of the 
autocorrelation function. The line shapes in the emergent spectrum
also depend on the distribution of absorbers as a function of depth,
which can be modelled with improved accuracy by new models of dust
condensation and settling. 

The far red K\,\textsc{i} wings of the latest T dwarfs still show
missing opacity in these models, a phenomenon similar to what has been
found for the Na\,\textsc{i} line profiles observed in extremely cool,
metal-rich white dwarfs. 
We show that the line profile in both cases 
is strongly determined by multiple-perturber interactions at short
distances and can no longer be reproduced by a density expansion, but
requires calculation of the full profile in a unified theory. 
Including such line profiles in stellar atmosphere codes will further
improve models for the coolest and densest dwarfs as well as for the
deeper atmosphere layers of substellar objects in general. 
\end{abstract}

\maketitle

%%%%%%%%%%%%%%%%%%%%%%%%%%%%%%%%%%%%%%%%%%%%
%% MAINMATTER
%%%%%%%%%%%%%%%%%%%%%%%%%%%%%%%%%%%%%%%%%%%%

\section{Introduction}

The alkali metals produce strong resonance absorption lines in all
late-type dwarfs. Competing opacity sources such as the molecular
bands of metal hydrides, titanium oxide or vanadium oxide, are removed
from the atmosphere as one proceeds to cooler effective temperatures,
and into the brown dwarf regime. 
First their opacity is replaced by continuous dust absorption
characteristic of the spectra of L dwarfs, but 
as in late L and T dwarfs condensate grains settle below their fully
radiative upper photosphere, the atmospheres become increasingly
clear. The  alkali elements bind less easily to molecules or grains,
and thus their ground state transitions remain the last optical
opacity sources, along with Rayleigh  scattering by H$_2$ and He. 
Due to the extreme transparency of the atmosphere their line wings
form under very high density conditions and thus show among the
strongest pressure broadening effects observed in stellar
atmospheres, due to collisions with H$_2$ and He. 
These massively broadened alkali lines therefore define
the local pseudo-continuum out to several thousands of
{\AA}ngstr{\"o}ms from the line cores of the K\,I and Na\,I\,D
doublets at 0.59 and 0.77~$\mu$m, as has been shown by
\citet{krlLdwarf99,burrowsMS00,LimDust}. 
The far wings of these lines show strong departures from Van der Waals
theory and a simple Lorentzian shape. Precise modelling of the line
shapes requires detailed quantum-mechanical calculations, 
using accurate inter-atomic potentials as demonstrated by
\citet{NicolePhysRev99,BurrVolNaK,Alkalis03}.  

\section{Brown dwarf model spectra}

Detailed line profiles for all alkali resonance doublets 
(Li, Na, K, Rb, Cs) have been calculated by 
\citet{Alkalis03,alkalisLi,NicoleRbCs06,Cores06,NicoleKH2sat} for
perturbation by both H$_2$ and He. 
These calculations, recently updated from the molecular  potentials of
\citet{Rossi85} for the H$_2$ interaction to the newer ones of
Spiegelman \citep[in preparation, see also][]{NicoleKH2sat}, are 
included in our stellar atmosphere models calculated with the
\texttt{PHOENIX} code \citep{hbjcam99}. 
Compared to the previous generation of brown dwarf spectra described
by \citet{LimDust}, the detailed profiles in particular of the
Na\textsc{i} and K\textsc{i} doublets have produced much improved
synthetic spectra of the red optical and near infrared region. 

\subsection{Alkali chemistry and condensation}
Since the core, near and far wings of these strong lines form along an
extended optical path through the atmosphere, one also 
needs to know the numbers of absorbing atoms at all depth points
contributing to the line. These can vary considerably with depth, as
alkali metals and other refractory elements are being depleted in
brown dwarf atmospheres by condensation onto grains, and eventually
sedimentation of the condensates into deeper layers
\citep{loddersAlkalis}.  
Condensation of dust particles has to be considered for effective
temperatures below 2500\,K, i.\,e.\ essentially for all dwarfs later
than spectral type M. At $T_{\rm eff} < 2000$\,K the top of 
the cloud deck starts to sink into deeper parts of the atmosphere, 
receding to the optically thick layers at $T_{\rm eff}$ of 
1200\,$\ldots$\,1400\,K, which marks the transition from spectral
class L to T. 
Condensate fractions for such objects therefore can no longer be
determined from chemical equilibrium calculations such as those used
in the equation of state of \citet{LimDust}. In the`Dusty' limit
described in the former work condensates would be assumed to be
present everywhere where they are thermodynamically stable, and to be
in condensation/evaporation equilibrium with the gas phase. 
In contrast, in our current \textbf{Settl} models the amount of dust
and the fraction of metals still present in the gas phase is
calculated by comparing timescales for grain growth and turbulent
mixing driven by convective overshoot. This approach has produced a
much more realistic cloud model and an improved description of the L/T
transition \citep{settl07}. 

\begin{figure}
  \includegraphics[height=.3\textheight,clip]{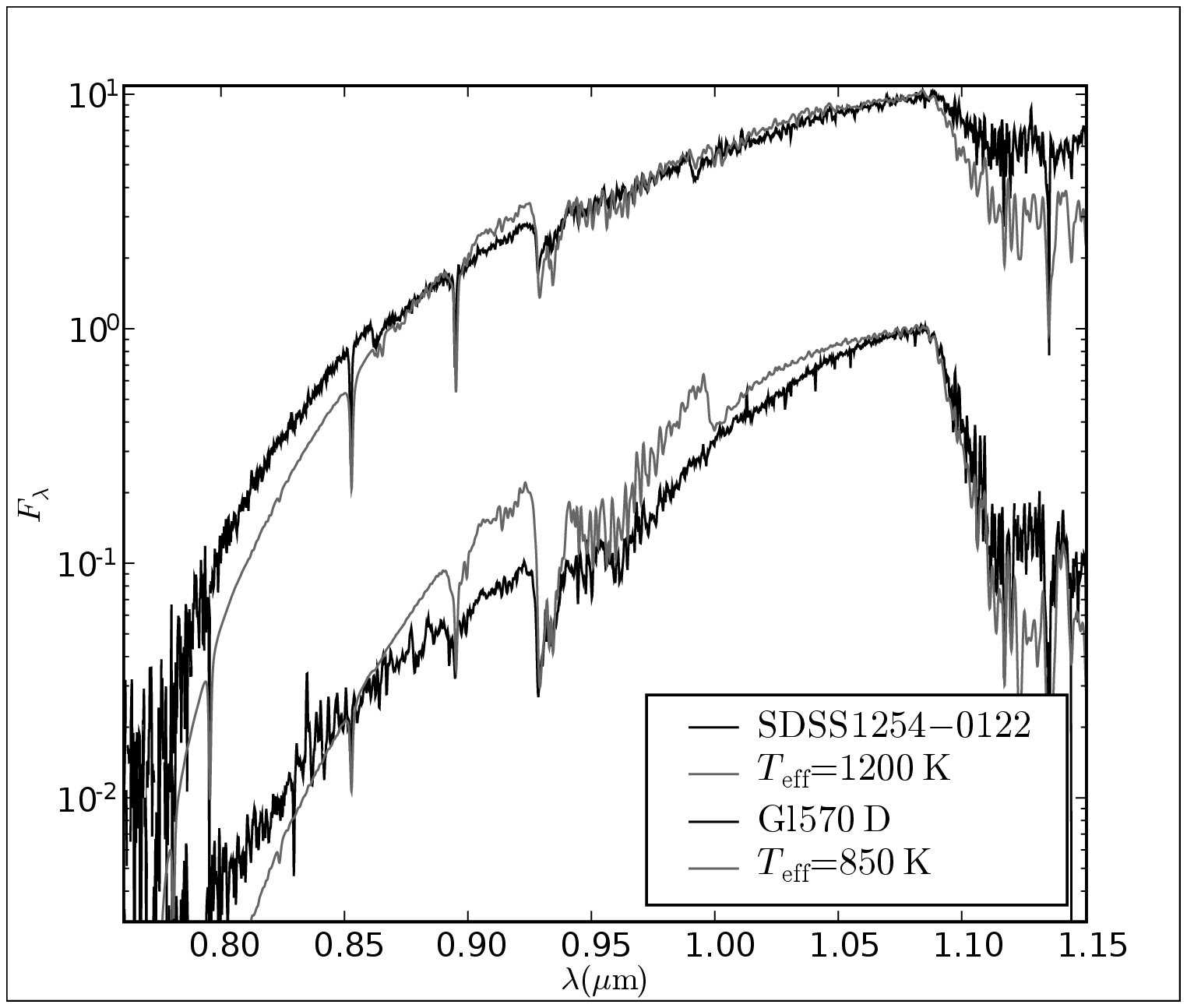}
  \includegraphics[height=.3\textheight,clip]{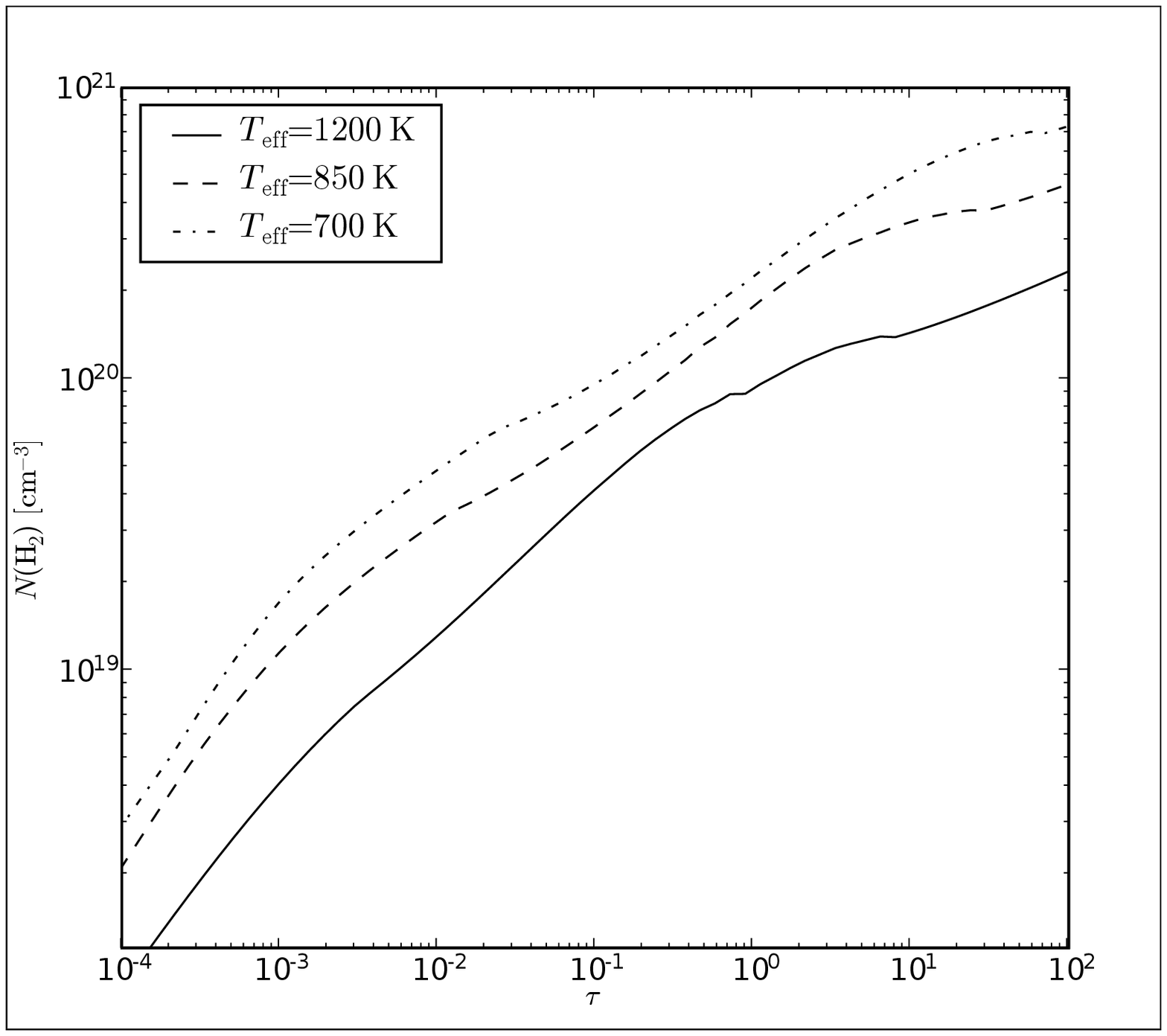} 
  \caption{Left: Red wing of the K\textsc{i} doublet, in the spectra of the
    T2 dwarf SDSS1254$-$0122 and the T8 dwarf Gliese~570\,D (black) 
    \citep{adam02a,adam03d}, 
    compared to \texttt{PHOENIX} models calculated for log\,$g$ of 5.25
    (gray). All spectra have been normalised to the peak of the $Y$-band
    flux peak at 1.08\,$\mu$m, and the T2 observation and model are
    shifted upward by a factor of 10. 
    Right: Number densities of molecular hydrogen as a function of
    optical depth for the atmosphere models shown on the left, and for
    a model for a very late T dwarf.
    \label{fig:tmodels}
  }
\end{figure}

\begin{figure}
  \includegraphics[height=.29\textheight,clip]{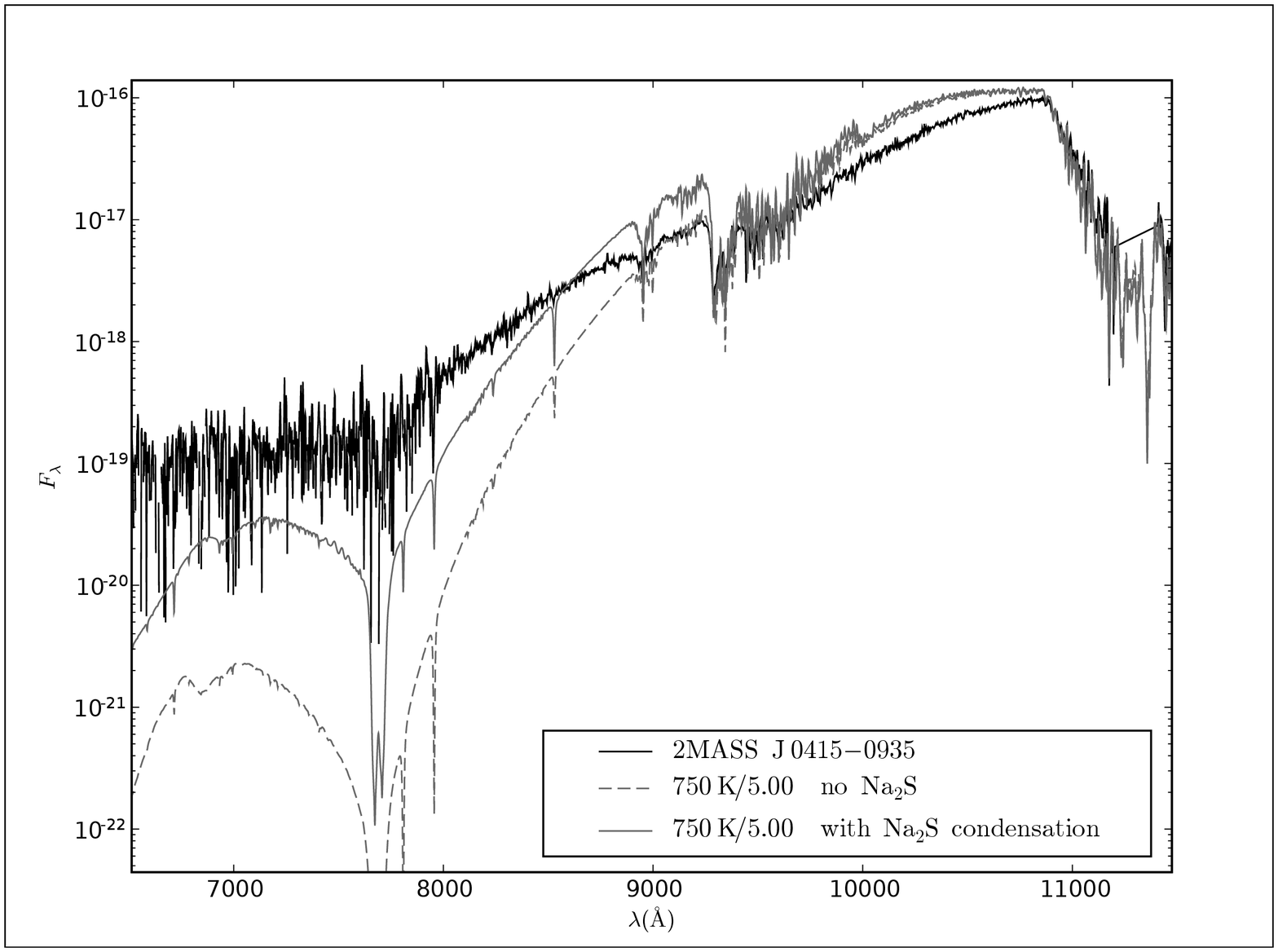}
  \includegraphics[height=.29\textheight,clip]{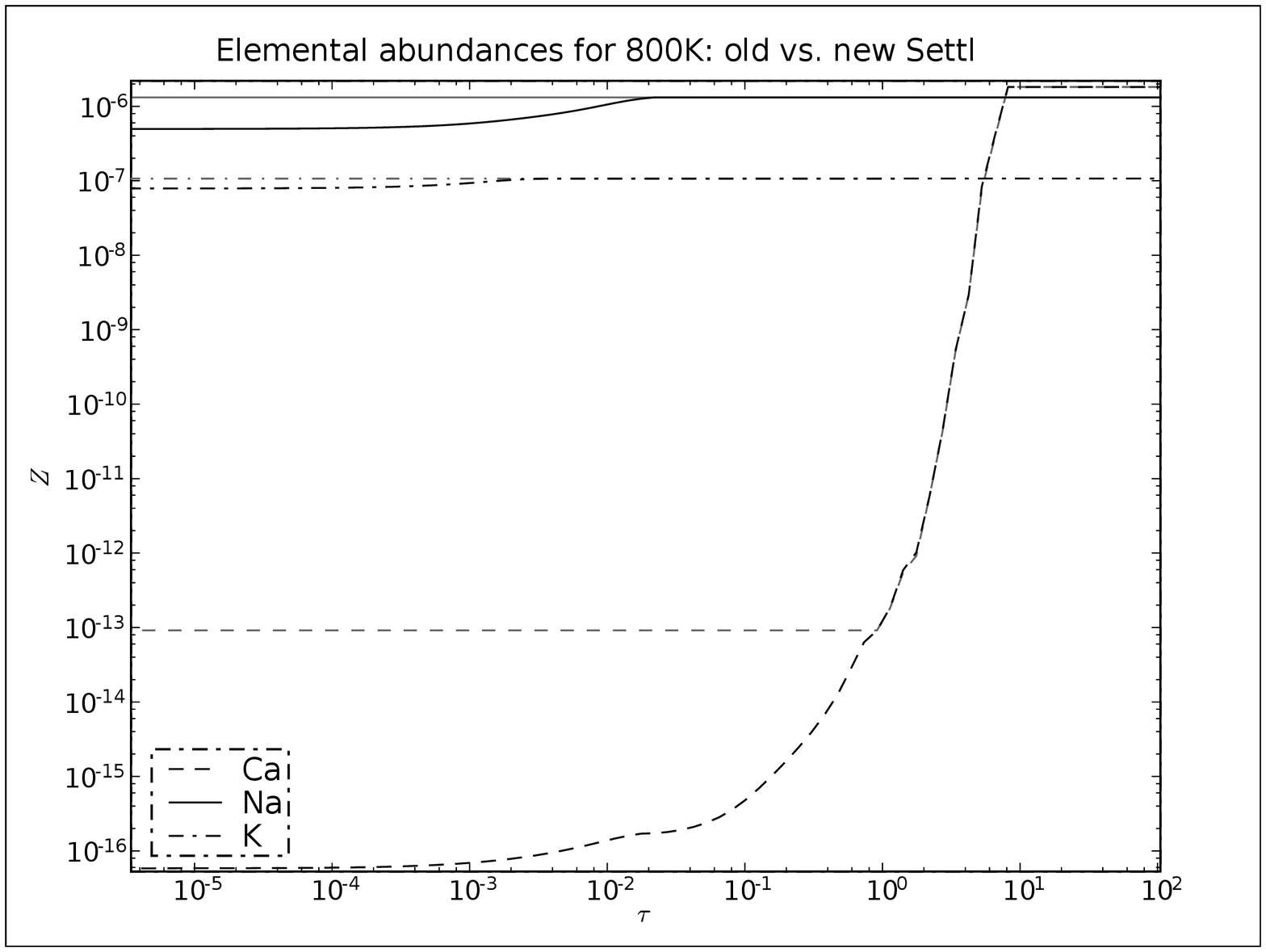}
  \caption{Left: Spectrum of the T8 dwarf 2MASS\,J0415$-$0935
    \citep{adam02a,adam03d} with \texttt{PHOENIX} Settl model
    including complete formation and sedimentation of grains (solid),
    and comparison with one alkali-bearing condensate left out of the
    equation of state (dashed). The latter not only inhibits
    condensation of sodium, but also leaves much higher concentrations
    of gaseous potassium in the atmosphere.
    Right: Abundances of important alkaline (earth) metals for partial
    (gray) and full (black) implementation of depletion
    effects.}
  \label{fig:2mmodels}
\end{figure}

The Settl models, combining the new line profile calculations with the
depth-dependent number densities of refractory elements, also show
very good agreement with the observed line shapes of alkali metals
(see Fig.~\ref{fig:tmodels}). Even for the L/T transition objects,
which are notoriously difficult to model, the optical spectrum is 
reproduced extremely well, allowing us to identify for the first time
the satellite feature located in the blue wing of the potassium line
\citep{satellites07}. 
But due to the complex cloud physics and chemistry results of the
settling model naturally depend sensitively on input parameters and
correct implementation. 
This is illustrated in Fig.~\ref{fig:2mmodels}, where the effect of
omitting just one of the species involved in the alkali metal
chemistry is shown. Removing one potential sink for sodium not merely
affects the fraction of sodium present in the gas phase, but also
those of the other alkali metals which are coupled to it by a chemical
reaction network, and produces strongly overestimated potassium
absorption. Such effects are most visible in the cores and near wings
of the alkali resonance lines, which form in the higher and cooler
layers that are most affected by the depletion processes. In these
regions the models still tend to predict too high concentrations of
K\,\textsc{i} and therefore too much absorption out to several
100\,{\AA} from the line core, indicating that we are still missing
some aspects of the cloud chemistry.  

\subsection{Limitations of the line profiles}
Comparison with the spectra of the coolest T dwarfs, such as Gl\,570D or
2MASS\,J0415$-$0935 shown in
Figs.~\ref{fig:tmodels}\&\ref{fig:2mmodels}, however also reveals
discrepancies in the very farthest parts of the K\,\textsc{i} line
wings, which extend to the red nearly 3000\,{\AA} into the $Y$-band
flux peak. While this region is still reproduced well in early and
mid-T dwarfs \citep[see also][]{satellites07}, the models
significantly underestimate the absorption in the far wing for late T
dwarfs. As this part of the spectrum forms very deep in the
atmosphere, where temperatures even in such cool brown dwarfs are too
high for any alkali condensation to occur, shortcomings of the cloud
model cannot be responsible for the mismatch in this case. 

A possible explanation may be found by testing the applicability of
our line formation code to the conditions of the coolest T dwarfs. 
We use profiles of the alkali metal resonance lines calculated for
perturbations due to neutral He and H$_2$  as the Fourier transform of
the autocorrelation function of the dipole moment within the adiabatic 
theory, as detailed in \citet{Alkalis03,alkalisLi}. 
In the \texttt{PHOENIX} implementation the depth-dependent line
opacity is calculated by splitting the profile into a core component,
which is describing the interactions at long distance in the impact
approximation, leading to a Lorentzian line shape, and the far wings
for close interactions that can produce the detunings of several
1000\,{\AA} observed in T dwarf spectra (see also the contribution
of G.~Peach, these proceedings). 
The latter is computed in the low density limit using an expansion of
the autocorrelation function in powers of density, which for
the present models has been developed to the third order and
evaluated at a perturber density of 10$^{19}$\,cm$^{-3}$. 
While this method allows an easy (linear) scaling to different
perturber densities, one has to be aware that only the core part
explicitly includes multiple perturber effects, while the non-linear
behaviour of the higher-order terms of the density expansion is not
correctly taken into account. 
In addition, \citet{alkalisLi} have shown that at higher densities,
where multiple perturber interactions become important even at close
distances, higher perturber density does not merely increase the line
strength, but affects the shape of the wings as well, and therefore
the low density limit is only strictly applicable at densities up to 
10$^{19}$\,cm$^{-3}$ in the case of Na or K broadened by H$_2$ or He,
and will quickly break down above 10$^{20}\,\mathrm{cm}^{-3}$ . 
The density profiles in Fig.~\ref{fig:tmodels} show that this limit
is quickly exceeded in T dwarfs at an optical depth of the order
unity, where the far red wing of the K\,\textsc{i} doublet
forms. While the H$_2$ density stays just below 10$^{20}$\,cm$^{-3}$
in mid-T dwarfs, it reaches several times that value for the latest T
types in the case of massive brown dwarfs. 
For atmospheres of less than the log\,$g=5.25$ shown here, densities
would be correspondingly lower, but such low gravities can only be
expected for rather young objects \citep{evolPlanets}.  
We thus find evidence that the K\,\textsc{i} absorption in the
$Y$-band is underestimated for the latest T dwarfs because their
atmospheres significantly exceed the density range where the present
line profile calculations may be applied. Line shapes calculated with
the unified theory of \citet{NicolePhysRev99} such as those described
by \citet{alkalisLi} should improve that situation. 

\begin{figure}
  \includegraphics[height=.28\textheight,clip]{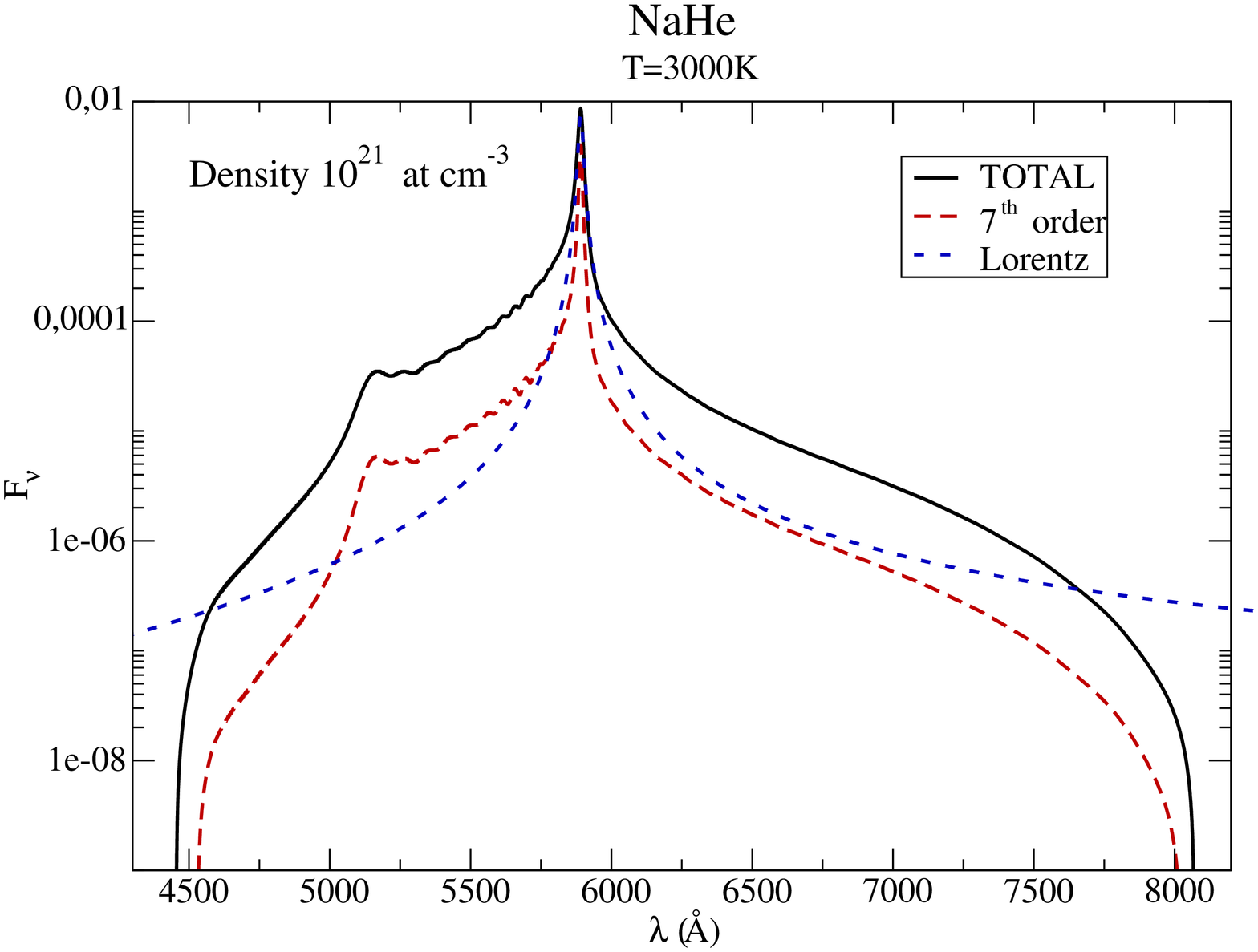}
  \includegraphics[height=.28\textheight,clip]{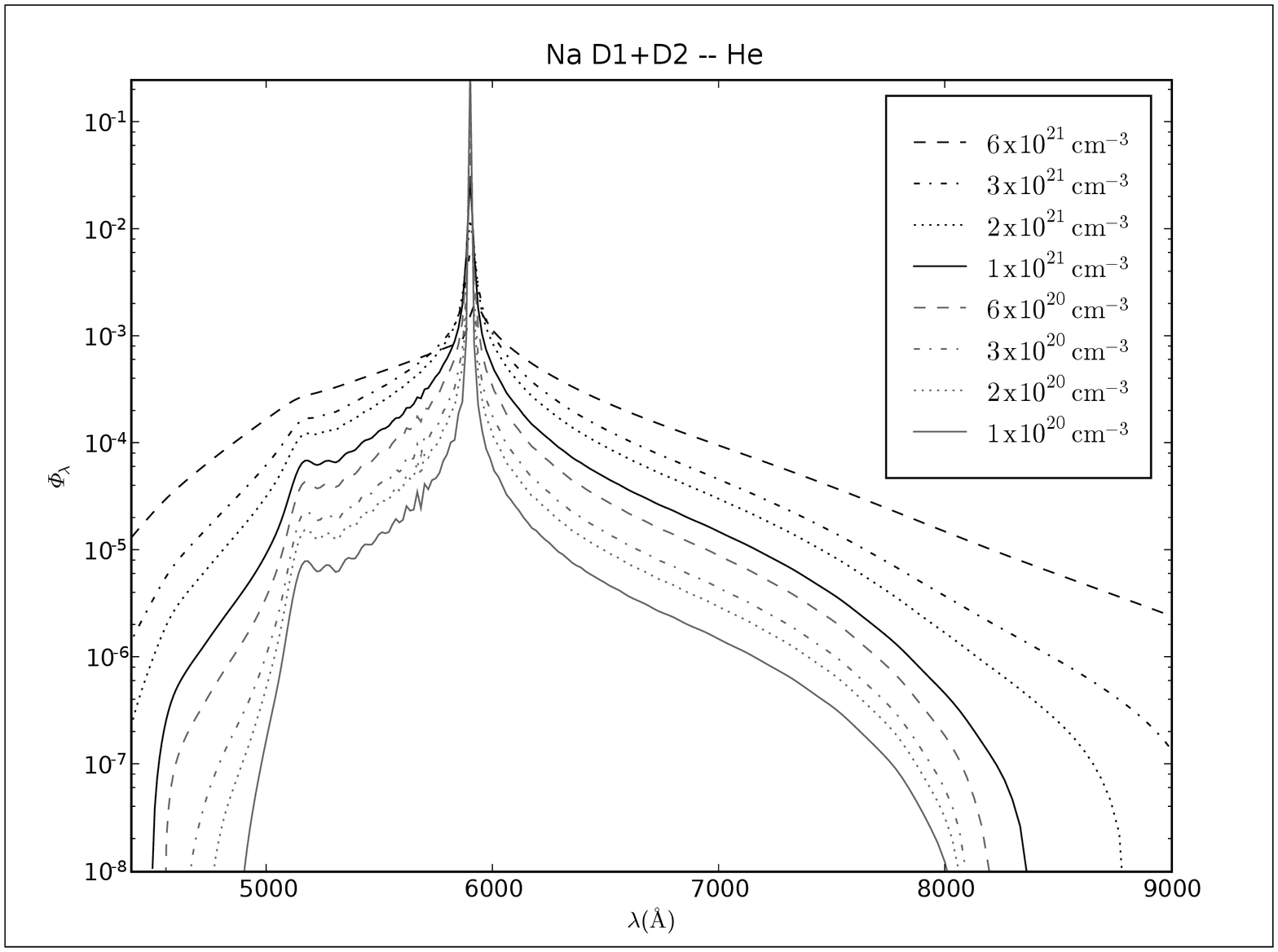}
  \caption{Left: Full unified profile of Na\,I\,D2 perturbed
    by 10$^{21}\,\mathrm{cm}^{-3}$ of He at 3000\,K, compared to
    the 7$^\mathrm{th}$ order expansion and Lorentzian
    \citep[from][]{alkaliWD15}.
    Right: Full unified profiles of the Na\,I~D2 line for He perturber
    densities from 1 to 60$\times$10$^{20}\,\mathrm{cm}^{-3}$, all 
    calculations at $T$\,$=$\,3000\,K.}
  \label{fig:NaHeProfs}
\end{figure}

\section{Alkali lines in cool white dwarfs}
\citet{alkaliWD15} have investigated possible broadening effects of
the Na\,\textsc{i}~D doublet at even higher perturber densities,
occuring for metal-rich white dwarfs with a helium-dominated
atmosphere. They found that the density of He as a perturber in two
very cool white dwarfs showing very strong Na absorption could reach
several 10$^{21}$ to 10$^{22}\,\mathrm{cm}^{-3}$, depending on exact
composition and temperature. 
They have directly compared the results of the density expansion model
and the unified theory for the Na\,\textsc{i}~D lines under these
conditions. 
The left side of Fig.~\ref{fig:NaHeProfs} shows that at a He 
density of 10$^{21}\,\mathrm{cm}^{-3}$ even carrying the density
expansion to the 7$^\mathrm{th}$ order rather than just to the
3$^\mathrm{rd}$, the line strength in the wings falls short of the
unified profile by almost an order of magnitude. 
Yet the impact approximation can only reproduce the profile within a
few 100\,{\AA} from the line core \citep[see also][]{Cores06}. 
The right hand side illustrates that at the highest densities the far
wings do not simply scale in strength, but also change their shape
towards a wider profile, especially above
10$^{21}\,\mathrm{cm}^{-3}$. 

We expect a similar behaviour for the line shapes of the K\,\textsc{i}
doublet with both He and H$_2$ as perturbers. While conditions in
brown dwarfs are somewhat less extreme than in ultracool white dwarfs,
this example shows that the coolest T dwarfs, too,
require line profile calculations taking multiple collision effects
into account in a unified theory. 

%Some url test \url{http://www.world.universe}.

\section{Discussion}
New line profiles have greatly improved spectral models of all the
alkali resonance lines. Remaining discrepancies with observed line
shapes of brown dwarfs for the cores and near wings can be traced to
shortcomings of the cloud model for the high atmosphere.  
A lack of absorption in the very far wings becomes evident in the
latest T dwarfs. Comparison with Na\,\textsc{i} lines observed in cool
white dwarfs support our 
interpretation that these discrepancies are due to the extreme
perturber densities, and that unified line profiles are needed to
model atmospheres at such high pressure. This will become more
important with cooler brown dwarfs still being discovered, and
spectral models for the yet to be found Y dwarfs needed. 
The additional opacity from alkali lines at high densities 
could also influence the radiative transfer in substellar atmosphere
models even below the visible photosphere, possibly affecting the
thermal structure, and thus cooling rate and evolution of both brown
dwarfs and giant gas planets.  
Metal-rich cool white dwarfs may provide an important testbed for
these new models. 

%%%%%%%%%%%%%%%%%%%%%%%%%%%%%%%%%%%%%%%%%%%%%%%%
%% BACKMATTER
%%%%%%%%%%%%%%%%%%%%%%%%%%%%%%%%%%%%%%%%%%%%%%%%

\begin{theacknowledgments}
We thank CINES, IDRIS and GWDG for providing computing resources to
this project. DH thanks 
everyone supporting his attendance at the VI.\ SCSLSA, especially Dragana
Ili{\'c} and Luca Popovi{\'c} from the LOC for providing last
minute logistic support, Sonja Schuh for filling in with teaching
responsibilities and Stefan Dreizler for funding. 
\end{theacknowledgments}



\bibliographystyle{aipproc}   % if natbib is available

\begin{thebibliography}{19}
\expandafter\ifx\csname natexlab\endcsname\relax\def\natexlab#1{#1}\fi
\providecommand{\enquote}[1]{``#1''}
\expandafter\ifx\csname url\endcsname\relax
  \def\url#1{\texttt{#1}}\fi
\expandafter\ifx\csname urlprefix\endcsname\relax\def\urlprefix{URL }\fi
\providecommand{\eprint}[2][]{\url{#2}}

\bibitem[{Kirkpatrick} et~al.(1999)]{krlLdwarf99}
J.~D. {Kirkpatrick}, I.~N. {Reid}, J.~{Liebert}, R.~M. {Cutri}, B.~{Nelson},
  C.~A. {Beichman}, C.~C. {Dahn}, D.~G. {Monet}, J.~E. {Gizis}, and M.~F.
  {Skrutskie}, \emph{ApJ} \textbf{519}, 802 (1999).

\bibitem[{Burrows} et~al.(2000)]{burrowsMS00}
A.~{Burrows}, M.~S. {Marley}, and C.~M. {Sharp}, \emph{ApJ} \textbf{531},
  438 (2000).

\bibitem[{Allard} et~al.(2001)]{LimDust}
F.~{Allard}, P.~{Hauschildt}, D.~{Alexander}, A.~{Tamanai}, and
  A.~{Schweitzer}, \emph{ApJ} \textbf{556}, 357 (2001).

\bibitem[{Allard} et~al.(1999)]{NicolePhysRev99}
N.~F. {Allard}, A.~{Royer}, J.~{Kielkopf}, and N.~{Feautrier}, \emph{{Phys.~Rev.~A}}
  \textbf{60}, 1021 (1999).

\bibitem[{Burrows} and {Volobuyev}(2003)]{BurrVolNaK}
A.~{Burrows}, and M.~{Volobuyev}, \emph{ApJ} \textbf{583}, 985 (2003).

\bibitem[{Allard} et~al.(2003)]{Alkalis03}
N.~F. {Allard}, F.~{Allard}, P.~H. {Hauschildt}, J.~F. {Kielkopf}, and
  L.~{Machin}, \emph{{A\&A}} \textbf{411}, L473 (2003).

\bibitem[{Allard} et~al.(2005)]{alkalisLi}
N.~F. {Allard}, F.~{Allard}, and J.~F. {Kielkopf}, \emph{{A\&A}} \textbf{440},
  1195 (2005).

\bibitem[{Allard} and {Spiegelman}(2006)]{NicoleRbCs06}
N.~F. {Allard}, and F.~{Spiegelman}, \emph{{A\&A}} \textbf{452}, 351 (2006).

\bibitem[{Allard} et~al.(2006)]{Cores06}
N.~F. {Allard}, F.~{Allard}, C.~{Johnas}, and J.~{Kielkopf},
\emph{submitted to {A\&A}} (2006). 

\bibitem[{Allard} et~al.(2007{\natexlab{a}})]{NicoleKH2sat}
N.~F. {Allard}, F.~{Spiegelman}, and J.~F. {Kielkopf}, \emph{{A\&A}}
  \textbf{465}, 1085 (2007{\natexlab{a}}).

\bibitem[{Rossi} and {Pascale}(1985)]{Rossi85}
F.~{Rossi}, and J.~{Pascale}, \emph{{Phys.~Rev.~A}} \textbf{32}, 2657 (1985).

\bibitem[Hauschildt and Baron(1999)]{hbjcam99}
P.~H. Hauschildt, and E.~Baron, \emph{J.~Comp.~Applied~Math.} \textbf{109}, 41
  (1999).

\bibitem[{Lodders}(1999)]{loddersAlkalis}
K.~{Lodders}, \emph{ApJ} \textbf{519}, 793 (1999).

\bibitem[{Allard} et~al.(2007{\natexlab{b}})]{settl07}
F.~{Allard}, D.~{Homeier}, T.~{Guillot}, G.~{Chabrier}, H.~{Ludwig},
  N.~{Allard}, C.~{Johnas}, J.~{Ferguson}, T.~{Barman}, and P.~{Hauschildt},
  \emph{{A\&A}} \emph{in prep.} (2007{\natexlab{b}}).

\bibitem[{Burgasser} et~al.(2002)]{adam02a}
A.~J. {Burgasser}, J.~D. {Kirkpatrick}, M.~E. {Brown}, I.~N. {Reid},
  A.~{Burrows}, J.~{Liebert}, K.~{Matthews}, J.~E. {Gizis}, C.~C. {Dahn}, D.~G.
  {Monet}, R.~M. {Cutri}, and M.~F. {Skrutskie}, \emph{ApJ} \textbf{564},
  421 (2002).

\bibitem[{Burgasser} et~al.(2003)]{adam03d}
A.~J. {Burgasser}, J.~D. {Kirkpatrick}, J.~{Liebert}, and A.~{Burrows},
  \emph{ApJ} \textbf{594}, 510 (2003).

\bibitem[{Allard} et~al.(2007{\natexlab{c}})]{satellites07}
F.~{Allard}, D.~{Homeier}, N.~{Allard}, M.~{McCaughrean}, F.~{Spiegelman},
  J.~{Kielkopf}, C.~{Johnas}, and P.~{Hauschildt}, 
  \emph{submitted to {A\&A}}, (2007{\natexlab{c}}).

\bibitem[{Baraffe} et~al.(2003)]{evolPlanets}
I.~{Baraffe}, G.~{Chabrier}, T.~S. {Barman}, F.~{Allard}, and P.~H.
  {Hauschildt}, \emph{{A\&A}} \textbf{402}, 701 (2003).

\bibitem[{Homeier} et~al.(2007)]{alkaliWD15}
D.~{Homeier}, N.~{Allard}, C.~{Johnas}, P.~{Hauschildt}, and F.~{Allard},
  in \emph{15th European Workshop on White Dwarfs},
  eds.\ R.~Napiwotzki, M.~Barstow {et~al.}, \emph{ASP Conf. Ser.}, San
  Francisco, \emph{in press} (2007).

\end{thebibliography}


\end{document}